# The Structure and Thermodynamic Stability of Reverse Micelles in Dry AOT/Alkane Mixtures

Adam Wootton, Francois Picavez and Peter Harrowell

*School of Chemistry, University of Sydney, Sydney, New South Wales 2006, Australia*

**Abstract.** Monte Carlo simulation studies of reverse micelles of an anionic surfactant, sodium AOT, in a non-polar solvent provide strong evidence that, in the absence of water, these clusters are charge ordered polyhedral shells. The stabilizing energy of these clusters is so large that the entropy of mixing is, in comparison, inconsequential and we predict that, if all waters of hydration could be removed (something not yet accomplished for the sodium salt) then AOT would be insoluble in nonpolar solvents.

## INTRODUCTION

The self assembly of amphiphiles in non-aqueous media represents an important and outstanding puzzle. In the absence of water, we cannot invoke the hydrophobic effects that govern aqueous self assembly into micelles. In water, these small aggregates represent a balance between two opposing fragile effects (i.e of the order of $k_BT$) : the entropy of mixing, on one hand, and the hydrophobic interaction between the surfactant's nonpolar 'tails' and the water, on the other. No such general explanation currently exists for reverse micelles (micelles with the non-polar groups on the exterior) that are observed in organic solvents. The best studied case is that of the anionic surfactant sodium bis (2-ethyl-hexyl) sulfosuccinate (commonly known as sodium Aerosol OT or sodium AOT).  Many surfactants are not oil-soluble, so compounds like sodium AOT represent a significant but select subset of amphiphiles, typically characterised by branched alkane chains. There has been considerable interest in using reverse micelles or, more specifically, the microemulsions obtained by swelling these reverse micelles with water, as microscopic reaction vessels. While there has been a number of theoretical studies of the reverse AOT micelles (summarised below), most have focussed on the properties of the confined water rather than the stability of the self assembled structure itself. The goal of this paper is to explore the energetics and thermodynamics of micelle formation in dry sodium AOT/alkane mixtures. In doing so we shall draw on our recent work in modeling the stability of ionic clusters with significant size asymmetry [1].

The experimental situation has been described in the comprehensive study by Ekwall, Mandell and Fontell [2]. Of particular significance we note that pure sodium AOT does not form a crystalline solid, rather it appears as an inverse hexagonal structure. Note that whereas the crystal structure has no connection with the normal micelle structure, it is, as we shall demonstrate, closely associated with that of the reverse micelle. In a sense, we shall argue, that reverse micelles are best regarded as representing a partial disruption of the crystalline structure. The forces that dominate crystallisation, hence, dominate reverse micelle formation. We are not aware of this point being made previously.

Salaniwal et al [3,4] have performed simulations on inverse microemulsions of sodium AOT and water ( $w \geq 10$) in supercritical $CO_2$. Senapati et al [5] also used $CO_2$ as a solvent but with phosphate fluoro-surfactants. Self assembly of roughly spherical micelles of the experimentally observed size was reported. The orientation and location of the confined waters were observed to be strongly influenced by the electric field of the ions.

Comprehensive MD simulations have also been performed to study the stability of micelles with respect to aggregation [6] and the optimum solvation shells of AOT molecules [7]. Both of these studies reported good agreement with experimental observations. Other studies have used MD studies on surfactants with $Ca^{2+}$ as the cations and $CO_3^{2-}$ as the ionic head group [8,9]. These highly charged ions were tightly bound and were reported essentially as solid cores as the ions were tightly bound.

Early simulations of reverse micelles used a simplified tail model in which the tails where represented as a soft repulsive sphere [10]. The tail has been represented by beads and springs [11-13] and flexible chains of soft sphere [14]. All studies emphasise that the key contribution of the tails is to prevent the head groups moving into the core of the micelle. It must be noted that these simulations were performed with w >> 1 and completely neglecting the tail would have permitted the complete hydration of the head group.

A number of models of reverse micelles have been reported in which the chain is discarded to be replaced by some constraint on the head group. In refs. [15-17] the head groups are constrained to lie on a spherical shell while the counterions and water are constrained to the interior. These simulations, for $1 \leq w \leq 10$, confirmed the distinction of bound and unbound waters. The artifacts arising from the restriction of the waters to the inside of the sphere was acknowledged in this work. Another variant pins the anions to the surface of the cluster and allows the cluster shape to vary while optimizing the positions of the counter ions. [18].

An early study by Kotlarchyk et al [19] attempted to fit the results of small angle neutron scattering (SANS) to a model. These workers noted that the centre of the micelle was going to be densely filled. They proposed that the head groups would form an icosahedron (the counter ions were ignored) with the water occupying the internal space.

## THERMODYNAMICS OF CLUSTERING IN AN IDEAL SOLUTION

We shall develop the theory of reverse micelle stability along similar lines to those followed in the theory of normal micelles in aqueous solution as presented by Ben-Shaul and Gelbart [20]

We shall consider reverse micelles containing N ionic surfactant molecules (with the associated counter ion bound). Assuming that dilute solution theory holds for the surfactant we can write the chemical potential of the N-cluster to be

$$\mu_N = \mu_N^{o,\rho} + k_B T \ln \rho_N \quad (1)$$

where $\mu_N^{o,\rho} = -k_B T \ln(q_N/V)$ with $q_N$ the partition function of the N-cluster and $\rho_N = n_N/V$, the number density of N-clusters. It is useful to refer concentrations back to the total amount of surfactant. To this end we define the mole fraction of surfactants to be found in the $n_N$ N-cluster as

$$X_N = Nn_N/(N_{tot} + Q) \quad (2)$$

where $N_{tot}$ and $Q$ are the total number of surfactant and organic solvent molecules respectively. Note that

$$\sum_N X_N = X \quad (3)$$

where $X = N_{tot}/(N_{tot}+Q)$ is the total mole fraction of surfactant present and represents the important experimental control parameter.

We now need to re-express the chemical potential in terms of $X_N$. We first note that

$$X_N = \frac{N\rho_N}{\rho_{tot}} \quad (4)$$

where $\rho_{tot} = N_{tot}/V$. Substituting, Eq. xx into Eq. xx gives

$$\mu_N = \mu_N^o + k_B T \ln(X_N/N) \quad (5)$$

where $\mu_N^o = \mu_N^{o,\rho} + k_B T \ln \rho_{tot}$. While strictly $\mu_N^o$ retains a dependence on the surfactant concentration through $\rho_{tot}$, for small concentrations $\rho_{tot} \approx \rho_{solvent}$ and, on this basis, we shall neglect this X dependence in the following analysis. At equilibrium we have

$$\mu_N = N\mu_1 \quad (6)$$

so that

$$\mu_N^o + k_B T \ln(X_N/N) = N(\mu_1^o + k_B T \ln X_1) \quad (7)$$

We can rearrange this relation to give the following expression for the cluster fraction $X_N$,

$$X_N/N = X_1^N \exp[\beta N(\mu_1^o - \mu_N^o/N)] \quad (8)$$

Following Israelachvili et al [21], we shall define the mean cluster size as

$$\overline{N} = \frac{\sum_{N>1} N^2 x_N}{\sum_{N>1} x_N} \qquad (9)$$

This average cluster size can also be written as

$$\overline{N} = \frac{\partial \ln(S - x_1)}{\partial \ln x_1} \qquad (10)$$

The variance $\sigma_N^2$ can be written as

$$\sigma_N^2 = \frac{\partial \overline{N}}{\partial \ln x_1} = \overline{N} \frac{\partial \overline{N}}{\partial \ln(S - x_1)} \qquad (11)$$

# THE CHEMICAL POTENTIAL OF THE SURFACTANT CLUSTER (WHERE WE TREAT DRY REVERSE MICELLES AS A PROBLEM OF IONIC CLUSTERING IN THE VACUUM).

To calculate the chemical potential difference $\Delta_N = \mu_N^o - N\mu_1^o$ between the cluster of N surfactants and N isolated surfactants we shall introduce three simplifications. While each sounds more monstrous than the last, we beg the reader's patience as we shall seek to justify these assumptions *a postori* on the basis of the energy scales involved. The three simplifications are: i) we shall replace the solvent by a vacuum, ii) neglect any contribution from the tails, and iii) model the sulfonate head groups as a spherical ion.

The neglect of the solvent can be relaxed with the use of the generalised Born model of Still and coworkers [22] but we do not expect any significant change to the major conclusions of this study. In this calculation we shall treat the tails of the surfactant as simply part of the nonpolar solvent. Our reasoning is that there would appear to be little change to the environment and degree of constraint of the chain from that of the isolated surfactant to that when the surfactant is incorporated on the surface of a small cluster with high curvature. (Neglect of the chains would become a problem as the curvature changed sign.) As described below, the surfactant tails are not required to constrain the AOT anions to the surface of the cluster, this being automatically accomplished (in the absence of water) by the size asymmetry of the ions. Probably the most serious of our simplifications is the neglect of the shape and charge distribution in the anionic head group.

From Eqs. 1 and 5 we have

$$\mu_N^o = -k_B T \ln(q_N / V) + k_B T \ln \rho_{tot} \qquad (12)$$

Following ref. [23], we can divide the aggregate partition function $q_N$ into the momenta and configurational contributions so that

$$q_N = \frac{8\pi^2 V \exp(-E_N / k_B T)}{\Lambda^{3N}} \qquad (13)$$

where the $8\pi^2$ represents the rotational configuration space of the aggregate (or surfactant) and $\Lambda = \left(\frac{h^2}{2\pi m k_B T}\right)^{1/2}$ is the thermal de Broglie wavelength with m being the mass of the individual ion pair. We have assumed that the ions in the aggregate bind so strongly that, at the temperatures of interest, we can neglect vibrations in the configurational contribution and assume that the aggregate structure is simply the minimum energy structure, with that minimum energy being $E_N$. We acknowledge a logical inconsistency here in that, strictly, we should not treat the vibrations classically in the case of such tight binding. We model the sulfonate head group as a soft spherical ion. The following interaction potential was used:

$$V_{ij} = \frac{\alpha}{r_{ij}^{12}} - \frac{\beta}{r_{ij}^6} + \frac{q_i q_j}{r_{ij}} \qquad (14)$$

where the charges on the sulfonate and sodium ions are q = -1 and +1, respectively, and $\alpha_{Na-Na}$ = 0.021025 kJmol$^{-1}$nm$^{12}$, $\alpha_{AOT-AOT}$ = 2.6341 kJmol$^{-1}$nm$^{12}$, $\beta_{Na-Na}$ = 0.7206 x 10$^{-4}$ kJmol$^{-1}$nm$^6$ and $\beta_{AOT-AOT}$ = 0.002617 kJmol$^{-1}$nm$^6$. The cross terms are obtained using $\alpha_{ij} = \sqrt{\alpha_{ii} \alpha_{jj}}$.

The ratio of the radius of the sulfonate ion over that of the sodium cation is 1.5 (taken as ($\alpha_{AOT-AOT}$/$\alpha_{Na-Na}$)$^{1/12}$). We have recently completed studies of clusters of ions with large size asymmetries [1]. Our main observation was that large size asymmetry (and cluster sizes with N < 120) results in energy minima corresponding to polyhedral shells rather than recognizable fragments of the bulk crystal phase. The relevance of this result in the context of reverse micelles is that the polyhedral shell automatically stabilizes the tail-bearing anions at the surface of the cluster without the need for the explicit inclusion of the tail or some equivalent constraint. We also note that the counter ions, rather than making up a dissociated double layer, form an integral part of the ordered cluster.

The size asymmetry of the sodium and sulfonate ions is similar to that of the AgI system that we have studied previously. The gallery of minimum energy clusters for the sodium is shown in Fig.1. Briefly, these minima represent the lowest energy structure obtained after applying a conjugate gradient minimisation to the potential energy to the complete set of possible clusters of size N consisting of trivalent vertices. This complete set was generated using an algorithm *CaGe* [24]. For clusters of N > 48 we have assumed that the lowest energy structure is a rod.

## ESTIMATING THE CMC AND AVERAGE CLUSTER SIZE

We shall begin by considering an ionic cluster of N ion pairs (each pair being a head group and counterion). It is worth pausing here to emphasise the essential conclusion from the previous consideration of aggregation energetics. In the absence of water, reverse AOT micelles are ionic clusters. In stark contrast to the case of normal micelles, the aliphatic chains serve no specific role in stabilising aggregation in nonpolar solvents. As we shall discuss in Section 5, the main role of the branched tails is to *destabilize* the crystal state and so enhance surfactant solubility in nonaqueous phases.

The average potential energy of the lowest energy ion clusters at T = 300K is well described by a fitting function provided by Eq.15. This form has been discussed previously [21] as arising from the contributions of the rod ends (*b*) and the 'barrel' (*Na*) and can be written as

$$\frac{E_N}{N} = \frac{b}{N} - a \qquad (15)$$

where a = 5.8 eV and b = 2.8 eV. Note 5.8eV corresponds, at 300K, to 200kT, a contribution well in excess of the contributions neglected such as the solvent polarizability and the chain entropy.

Based on the estimates described above, we shall approximate the chemical potential $\mu_N^o$ for the N-cluster as

$$\mu_N^o \approx \frac{E_N}{N} \qquad (16)$$

Israelachvili et al [21] have considered the equilibrium distribution of micelles for the case when the chemical potential for the N-cluster can be written as $\mu_\infty^o + b/N$ (i.e similar to our Eq.15). In this case one can write

$$\frac{X_N}{N} = e^{-\alpha} Y^N \qquad (17)$$

where α = b/kT and

$$Y = X_1 \exp[(\mu_1^o - \mu_\infty^o)/kT] \qquad (18)$$

The total amphiphile concentration S can be written as

$$S = X_1 + e^{-\alpha} \sum_{N=2}^{\infty} N Y^N \qquad (19)$$

The sum in Eq. 19 can be performed analytically to give

$$S = X_1 + \frac{2Y^2 e^{-\alpha}}{1-Y}\left(1 + \frac{Y}{2(1-Y)}\right) \qquad (20)$$

Using Eqs.10 and 19 we have the following expression for the average cluster size

$$\overline{N} = 2 + \frac{Y}{1-Y}\left(1 + \frac{1}{2(1-Y)+Y}\right) \qquad (21)$$

Given a value of S, the total concentration of amphiphile, and explicit expressions for the cluster chemical potentials, Eq.20 provides, in principle, as explicit solution for $X_1$, the concentration of free surfactant. In the case where the energetics favours large clusters, we can obtain an approximate analytical expression for $X_1$. From Eq.17 we can see that for $X_N$ to be significant at large N, Y ≈ 1. In this case we can approximate S and $\overline{N}$ by

$$S \approx X_1 + \frac{e^{-\alpha}}{(1-Y)^2} \quad (22)$$

and

$$\bar{N} \approx \frac{2}{1-Y} \quad (23)$$

From Eq.22 we have

$$X_1 \approx (1-(Se^\alpha)^{-1/2})\exp\left[(\mu_\infty^o - \mu_1^o)/kT\right] \quad (24)$$

For all $S > 10^3$ $e^{-\alpha}$, this gives us $X_1 \approx \exp\left[(\mu_\infty^o - \mu_1^o)/kT\right]$. This result is consistent with our assumption that $Y \approx 1$. More importantly, it predicts that the surfactant is insoluble at any accessible temperature. Consideration of the average cluster size underscores this conclusion. We can use Eq.22 to eliminate Y in Eq.23 to give

$$\bar{N} \approx 2\sqrt{Se^\alpha} \quad (25)$$

Taking solutions of $10^{-4}$ % and 1% by weight of sodium AOT in decanol, Eq. 25 gives an estimate of the number of surfactants per cluster of $10^{21}$ and $10^{22}$, respectively. Such large rods would certainly aggregate and fall out of solution. Clearly the large ionic stabilization energy precludes solubility of such ionic surfactants in a nonpolar medium.

## CONCLUSIONS

The AOT clusters are ordered along the tubes and insoluble in organic solvents. Should it be possible to produce a hexagonal solid consisting of a single domain of aligned tubes then we predict that x-ray scattering would observe the periodicity of the ion ordering. The tubes which pack to form the hexagonal solid phase are stabilised by the asymmetry in ion size. This purely ionic effect acts quite independently of any additional stabilization of the tubes arising from excluded volume effects of the chains.

To our knowledge sodium AOT has not been dried to less than 0.7 water molecules per surfactant. We conclude that this trace water is essential for AOT solubility and must act to break up the extended tubes. Preliminary Monte Carlo simulations of the ionic clusters studied here in the presence of a small amount of water strongly supports this proposition. We examine the role of waters of hydration n stabilizing reverse AOT micelles in a paper currently in preparation.

The role of chains in reverse micelles is quite different from their role in aqueous micelles. In reverse micelles they function only to disrupt the solid phase. This means that, whereas in normal micelles the degree of branching of the alkane chains is of relatively minor significance with respect to micelle stability, such branching is all important in micelles in organic solvents.

## ACKNOWLEDGMENTS

We would like to thank Greg Warr for valuable and timely guidance. AW acknowledges the support of the Gritton and Joan Clark scholarships.

## REFERENCES

1. A.Wootton and P.Harrowell, *J.Chem.Phys* **121**, 7440-7442 (2004); *J.Chem.Phys.B* **108**, 8412-8418 (2004).
2. P. Eckwall, L. Mandell and K. Fontell, *J. Coll. Int. Science* **33**, 215-235 (1970).
3. S.Salaniwal, S.T.Cui, H.D.Cochran and P.T.Cummins, *Langmuir*, **17**, 1784-1792 (2001)
4. S.Saliniwal, S.K.Kumar and A.Z.Panagiotopoulos, *Langmuir* **19**, 5164-5168 (2003)
5. S.Senapati and M.Berkowitz, *J.Chem.Phys.* **118**, 1937-1944 (2003)
6. M. Alaimo and T. Kumosinski, *Langmuir* **13**, 2007-2018 (1997)
7. B.Derecskei, A. Derecskei-Kovacs and Z. Schelly, *Langmuir* **15**, 1981-1992 (1999)
8. J.Griffiths and D.Heyes, *Langmuir* **12**, 2418-2424 (1996)
9. D.Tobias and M.Klein, *J.Phys.Chem.* **100**, 6637-6648 (1996)
10. D.Brown and J.Clarke, *J.Phys.Chem.* **92**, 2881-2888 (1988)
11. B.Smit, P.Hilbers, K.Esselink, L.Rupert, N. Van Os and A.Schlijper, *Nature* **348**, 624-625 (1990)
12. B.Smit, P.Hilbers, K.Esselink, L.Rupert, N. Van Os and A.Schlijper, *J.Phys.Chem.* **95**, 6361-6368 (1991).
13. B.Smit, K.Esselink, P.Hilbers, P.Van Os, l.Rupert and I.Szleifer, *Langmuir* **9**, 9-11 (1993)
14. S.Karaboni and R. O'Connell, *Langmuir* **6**, 905-911 (1990)
15. J.Faeder and B.Ladanyi, *J.Phys.Chem.B* **104**, 1033-1046 (2000)
16. J.Faeder, M.Albert and B.Ladanyi, *Langmuir* **19**, 2514-2520 (2003)
17. J.Faeder and B.Ladanyi, *J.Phys.Chem.B* **105**, 11148-11158 (2001)
18. A.Bulavchenko, A.Bastishchev, E.Batishcheva and V.Torgov, *J.Phys.Chem B* **106**, 6381-6389 (2002)

19. M.Kotlarchyk, J.Huang and S.Chen, *J.Phys.Chem*. **89**, 4382-4386 (1985).
20. A. Ben-Shaul and W. M. Gelbart, in *Micelles, Membranes, Microemulsions and Monolayers,* ed. A. Ben-Shaul, W. Gelbart and D. Roux, Berlin Spinger, 1994, p.1-28.
21. J. N. Israelachvili, D. J. Mitchell and B. W. Ninham, *J. Chem. Soc. Faraday Trans. 2* **72**, 1525-1530 (1976)
22. W.C.Still, A.Tempczyk, R.Hawley and T.Hendrickson, *J.Am.Chem.Soc*. **112**, 6127-6131 (1990).
23. R.Nagarajan and E.Ruckenstein, *Langmuir* **7**, 2934-2969 (1991)
24. The CaGe software was developed in the Department of Mathematics at the University of Bielefeld. At the time of writing; the software is available from http://www.mathematik.uni-bielefeld.de/CaGe/Archive/.

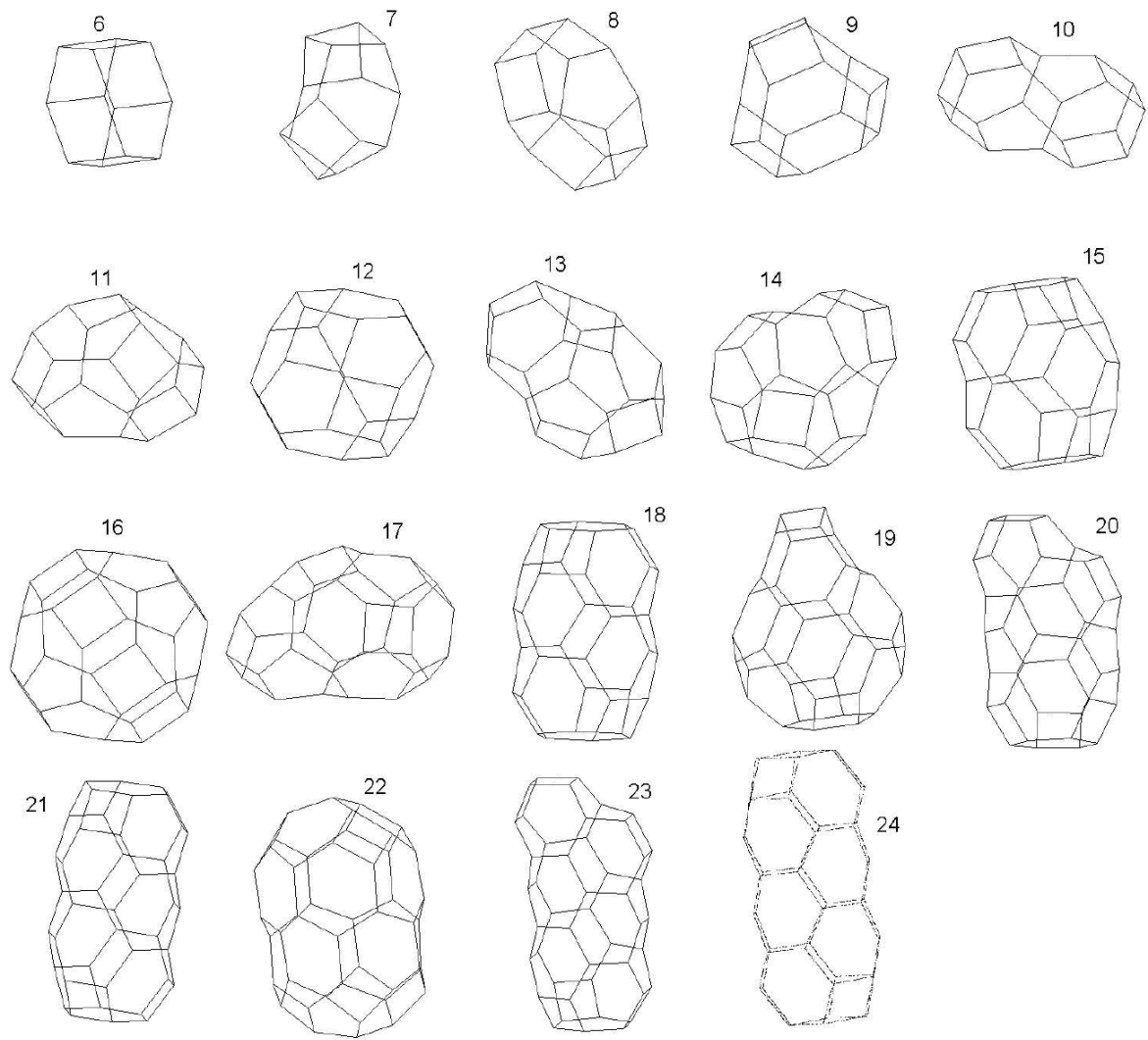

**FIGURE 1**. Structures of the lowest energy clusters for the sodium-sulfonate system as described in the text. N refers to the number of surfactants or, equivalently, the number of anion-cation pairs.